\documentclass[aps,prx,twocolumn,superscriptaddress,nofootinbib,longbibliography]{revtex4-2}

\pdfoutput=1  

\usepackage{graphicx}
\usepackage{amsmath,amssymb,amsfonts}
\usepackage[colorlinks,citecolor=blue,linkcolor=blue,urlcolor=blue, breaklinks=true]{hyperref}
\usepackage{color}
\usepackage[normalem]{ulem}

\usepackage[utf8]{inputenc}
\usepackage[english]{babel}
\usepackage[T1]{fontenc}

\usepackage{array,epsfig}

\usepackage{amsxtra}
\usepackage{amsthm}
\usepackage{mathrsfs}
\usepackage{color}
\usepackage{mathtools}
\usepackage{simplewick}
\usepackage{stmaryrd}
\usepackage{cancel}

\usepackage{newtxtext}
\usepackage[varg]{newtxmath}
\usepackage{csquotes}
\usepackage{appendix}

\usepackage{physics}
\usepackage{bm}

\usepackage{xcolor}
\usepackage{xspace}
\usepackage{ifthen}
\usepackage[normalem]{ulem}

\begin{document}


\title{Evidencing Quantum Gravity with Thermodynamical Observables}

\begin{abstract}

Proposed experiments for obtaining empirical evidence for a quantum description of gravity  in a table-top setting focus on detecting quantum information signatures, such as entanglement  or non-Gaussianity production, in gravitationally interacting quantum systems. Here, we explore an alternative approach where the quantization of gravity could  be inferred through measurements of macroscopic, thermodynamical quantities, without the need for addressability of individual quantum systems.  To demonstrate the idea, we take as a case study a gravitationally self-interacting Bose gas, and consider its heat capacity. We find a clear-cut distinction between the predictions of a classical gravitational interaction and a quantum gravitational interaction in the heat capacity of the Bose gas.
\end{abstract}


\author{Thomas Strasser}
\affiliation{Institute for Quantum Optics and Quantum Information (IQOQI) Vienna, Austrian Academy of Sciences, Boltzmanngasse 3, A-1090 Vienna, Austria}

\author{Marios Christodoulou}
\affiliation{Institute for Quantum Optics and Quantum Information (IQOQI) Vienna, Austrian Academy of Sciences, Boltzmanngasse 3, A-1090 Vienna, Austria}

\author{Richard Howl}
\affiliation{Department of Physics, Royal Holloway, University of London, Egham, Surrey, TW20 0EX, United Kingdom}

\author{\v Caslav Brukner}
\affiliation{Institute for Quantum Optics and Quantum Information (IQOQI) Vienna, Austrian Academy of Sciences, Boltzmanngasse 3, A-1090 Vienna, Austria}
\affiliation{Vienna Center for Quantum Science and Technology (VCQ), Faculty of Physics, University of Vienna, Boltzmanngasse 5, A-1090 Vienna, Austria}

\maketitle

Proposals for table-top tests of quantum gravity typically involve preparing a sufficiently massive system in a quantum state and then analysing  its quantum information properties to evidence non-classicality of states of its gravitational field.  Examples include seeking entanglement between two massive systems prepared in Schrödinger cat states \cite{PhysRevLett.119.240401, Marletto:2017kzi} and detecting quantum non-Gaussianity in a squeezed Bose-Einstein condensate \cite{Howl:2020isj}. This general  methodology presents a substantial technical challenge due to the need for preparing and isolating the system from the environment and detecting such highly non-classical quantum states.

Here, we explore whether an alternative approach might be possible. Inspired by ideas of verifying entanglement via macroscopic thermodynamic quantities, such as magnetic susceptibility and heat capacity \cite{Wie_niak_2005, brukner2004macroscopicthermodynamicalwitnessesquantum, Brukner_2006, Wie_niak_2008, Singh_2013}, we consider whether a quantum gravitational effect could  be detected through thermodynamic observables, thereby avoiding the need for coherent control or  the application of quantum operations and measurements. 

As a case study, we consider the self-gravitation contributions to the  heat capacity of a Bose-Einstein condensate (BEC). Assuming a classical linearized gravitational field  in the spirit of `semi--classical gravity' \cite{ROSENFELD1963353,moller1962theories}, where the self--gravitational field of the BEC is treated classically and given by the expectation of the quantum state of the BEC, we find that the energy spectrum and the heat capacity remain unaffected by the self--gravity interaction in the Newtonian regime. Intuitively, this is due to the interaction being `averaged' over the state of the BEC, which results in  the energy spectrum simply being offset by a constant, and has no effect on the heat capacity as it only depends on the spacing of the energy eigenvalues. In contrast, in  linearized quantum gravity, where the gravitational field of the BEC is treated quantum mechanically, correlations are induced between the atoms that cause a non-trivial modification of the energy spectrum in the Newtonian regime, resulting in a dependence of the  heat capacity on Newton's constant. Moreover, since the gravitational field cannot be shielded, the self-gravity contribution to the energies and the heat capacity is not restricted to the neighbouring atoms as in the case of the electromagnetic interaction, but extends to all atoms in the sample. The measurement of heat capacity or internal energy can, therefore,  provide a clear-cut criterion for distinguishing semi--classical gravity and linearized quantum gravity.

\medskip

\emph{Setup --- }The   grand--canonical Hamiltonian of a Bose gas with two-body electromagnetic interactions  reads
\begin{equation} \label{eq:K}
     \hat{K} = \int \!\!\! d^3 \mathbf{r} \; \hat{\Psi}^{\dagger} (\mathbf{r}) \left( - \dfrac{\hbar^2 \nabla^2}{2 m} - \mu \right) \hat{\Psi} (\mathbf{r})  + \hat{H}^{EM}_I, 
\end{equation}
where $\mu$ denotes the chemical potential,  $\hat{\Psi}^{\dagger}(\bm{r})$ is the non-relativistic quantum field of  the Bose gas, the first term represents the kinetic  energy of the gas, and $\hat{H}^{EM}_{I}$ is due to the electromagnetic interactions:  
\begin{equation}
\label{eq:HI}
    \hat{H}^{EM}_{I} = \frac{1}{2}\, \!\! \int \!\!\! d^3 \mathbf{r'}  d^3 \mathbf{r} \;  \hat{\Psi}^{\dagger} (\mathbf{r}) \hat{\Psi}^{\dagger} (\mathbf{r'}) V_I(\bm{r} - \bm{r'}) \hat{\Psi} (\mathbf{r'}) \hat{\Psi} (\mathbf{r}),
\end{equation}
with $V_I(\bm{r}-\bm{r'})$ the two-body electromagnetic potential. For simplicity, we  consider a Bose gas trapped in a box of length $L$ with periodic boundary conditions for the field, see for instance experiments \cite{UniformBECExp,PhysRevLett.118.210401,PhysRevLett.119.190404,PhysRevLett.119.250404} and theoretical work \cite{PitaevskiiBook}.

For a dilute Bose gas made up of neutral atoms, the electromagnetic interactions fall off significantly faster than $1/r$.  It is, therefore, a good approximation to treat the electromagnetic self--interaction as a contact interaction $V_I (\bm{r}) = g^{EM} \, \delta^{(3)} (\bm{r})$,  where $g^{EM} = 4 \pi \hbar^2 a_s / m $, with $m$ the particle mass and $a_s$ the s-wave scattering length. This is the typical treatment found in textbooks.\footnote{For BECs with non-contact electromagnetic interactions, such as electromagnetically charged BECs, see e.g., \cite{2014PhRvA..90f3636G,LUKIN2021127695}.} However, since gravity cannot be shielded, to study the gravitational self--interaction of the Bose gas it is crucial to keep explicit the long-range nature of gravity.

We will study two different types of gravity self--interaction terms. These correspond to a linearized quantized metric perturbation and to a classical gravitational field sourced by quantum matter, the BEC. In the non-relativistic limit of linearized quantum gravity (meaning zeroth order in the speed of light, but $G$ and $\hbar$ are relevant), the interaction term due to gravity is of the form \eqref{eq:HI} with $V_I(\bm{r}) = - G m^2 / |\bm{r}|$,
\begin{equation}
\label{eq:HIG}
    \hat{H}_{I}^{QG} = - \frac{1}{2}\,G m^2 \!\! \int \!\!\! d^3 \mathbf{r'} \!\! \int \!\!\! d^3 \mathbf{r} \;  \hat{\Psi}^{\dagger} (\mathbf{r}) \hat{\Psi}^{\dagger} (\mathbf{r'}) \frac{1} { |\bm{r-r'}|} \hat{\Psi} (\mathbf{r'}) \hat{\Psi} (\mathbf{r}),
\end{equation}
which is derived from first-principles in the Supplementary Information. Considering a classical metric perturbation in the spirit of `semi--classical gravity' \cite{ROSENFELD1963353,moller1962theories}, where gravity is sourced by the expectation of quantum matter, the interaction Hamiltonian is of the form
\begin{equation} \label{eq:HICG}
    \hat{H}^{CG}_{I} = - Gm^2 \int \!\! d^3 \mathbf{r'} \int \!\! d^3 \mathbf{r} \, \, \dfrac{\langle  \hat{\Psi}^{\dagger} (\mathbf{r'}) \hat{\Psi}(\mathbf{r'}) \rangle }{\vert \mathbf{r} - \mathbf{r'} \vert }  \, \hat{\Psi}^{\dagger} (\mathbf{r}) \hat{\Psi} (\mathbf{r}),
\end{equation}
 where the expectation value is taken over the quantum state of the Bose gas \footnote{There is an interesting connection between this mean-field gravity approximation and a mean-field approximation often used in describing BECs, which we discuss in the Supplementary Material.}. The reason the two cases of self gravitation will give very different contributions to the heat capacity is that $\hat{H}^{CG}_{I}$ is of \emph{quadratic} order in the annihilation and creation operators, while $\hat{H}_{I}^{QG}$ is of \emph{quartic} order. 

\emph{Bogolyubov method applied to long range interaction ---} The Bogolyubov method is the standard method to analyse a homogeneous, isotropic gaseous BEC  with contact electromagnetic interactions \cite{PitaevskiiBook}, which is the type of BEC we assume for simplicity, see for instance experiments \cite{UniformBECExp,PhysRevLett.118.210401,PhysRevLett.119.190404,PhysRevLett.119.250404}. We adapt and apply this method to a $1/r$ (long range) interaction term, with the aim to arrive at a diagonalized Hamiltonian from which the heat capacity can be obtained. The Bogolyubov method consists principally of three major steps: 1) The Bogolyubov approximation, 2) A Fourier transform to momentum space, and 3) A linear symplectic transformation (Bogolyubov transformation) that diagonalizes the Hamiltonian. Finally, we use the gapless condition  to determine the chemical potential  \cite{2006PhRvA..73f3612Y}.

\emph{Bogolyubov approximation --- }The first  step is to split the field operator as $\hat{\Psi}(\mathbf{r}) = \hat{\phi}(\mathbf{r}) + \hat{\psi}(\mathbf{r})$, where $\hat{\phi}(\mathbf{r})$ and $\hat{\psi}(\mathbf{r})$ are respectively for the condensate and excited states.  Taking the definition of a BEC to be the macroscopic occupation of the ground state, i.e.\ $N \gg 1 $ and $N \approx N_0$, where $N_0$ and $N$ are the total number of atoms and number of condensed atoms respectively, the Bogolyubov approximation is that the field operator $\hat{\phi}(\mathbf{r})$  is  traded for a classical variable $\phi$ using the substitution $\hat{\phi}(\mathbf{r})=\sqrt{n}$, where $n := N / \mathcal{V}$ is the number density of the Bose gas, with  $\mathcal{V} = L^3$  the volume of the box.\footnote{Note, as stated above, we assume the BEC to be uniform, and thus we do not consider a Gross-Pitaevskii equation for the condensate (see e.g.\ \cite{PitaevskiiBook}). This is discussed further in the Supplementary Material.}  Inserting the above decomposition into the full grand-canonical Hamiltonian $\hat{K}$, only terms with two or more  $\hat{\phi}(\mathbf{r})$ are kept since $N \gg 1$ \cite{PitaevskiiBook}. All other terms are taken negligible. An explicit derivation is provided in the Supplementary Material.

\emph{Fourier transform ---} Next, we Fourier expand the excitation fields as
    $\hat{\psi} (\bm{r}) = \frac{1}{\sqrt{\mathcal{V}}} \sum_{\bm{k}}    e^{i\bm{k} \cdot \bm{r}} \hat{a}_{\bm{k}} $,
where  $\hat{a}_{\bm{k}}$ is the annihilation  operator for mode $\bm{k}$. Similarly,  the interactions are expanded as
$V_I(\bm{r}) = \frac{1}{\mathcal{V}} \sum_{\bm{k}} V_{\bm{k}} e^{i \bm{k}\cdot\bm{r}}$, where
$V_{\bm{k}} =  \int_\mathcal{V} d^3 \bm{r} \, V_I(\bm{r})\, e^{-i \bm{k}\cdot\bm{r}}$,
with $\bm{k} = 2 \pi \bm{n}_k / L$,  $\bm{n}_k = (n_x, n_y, n_z)$, and $n_x,~n_y,~n_z$ integers ranging from $-\infty$ to $+\infty$. 

For the electromagnetic interactions, which are modelled as contact interactions $V_I(\bm{r}) = g^{EM} \delta^{(3)}(\bm{r})$, we simply have $V_{\bm{k}} = g^{EM}$ for all momenta $\bm{k}$. For the long-range gravitational interaction, we must calculate 
\begin{align} \label{eq:VkCube}
    V^{G}_{\bm{k}} &=  \int d^3 \bm{r} \frac{- G m^2 }{|\bm{r}|} e^{-i \bm{k}\cdot\bm{r}}\\
    &= \int^{\frac{L}{2}}_{-\frac{L}{2}} dx \int^{\frac{L}{2}}_{-\frac{L}{2}} dy \int^{\frac{L}{2}}_{-\frac{L}{2}} dz   \frac{- G m^2 e^{-2 \pi i  (n_x x + n_y y + n_z z)/L}  }{\sqrt{x^2 + y^2 +z^2} }.
\end{align}
 This is well approximated by $V^{G}_{\bm{k}}=g^G_{\bm{k}}$ where
\begin{align}\label{eq:Vq}
    g^G_{\bm{k}}  = \begin{cases}
   g^G_0  & k = 0  \\
   -4 \pi G m^2/k^2  & k > 0,
\end{cases}
\end{align}
with $g^G_0 \approx L^3 V_{cube}$, and  $V_{cube} = - 2.38 G m^2 / L$  the  gravitational potential of a cube of side $L$ at its centre. A derivation is given in the Supplementary Material.

Inserting  the above steps into  the quantized gravity interaction Hamiltonian \eqref{eq:HIG}, we obtain
\begin{align} \label{eq:HIQG}
     \hat{H}_I^{QG} = \frac{ 1}{2} n \left(  N g^G_0 + \sum_{\bm{k} \neq \bm{0}}   g^G_k  \hat{A}_{\bm{k}} \right),
\end{align}
with $\hat{A}_{\bm{k}} = 2 \hat{a}^{\dagger}_{\bm{k}} \hat{a}_{\bm{k}} + \hat{a}^{\dagger}_{\bm{k}} \hat{a}^{\dagger}_{-\bm{k}}  +  \hat{a}_{\bm{k}} \hat{a}_{-\bm{k}} $. Compare this to the electromagnetic interaction Hamiltonian, which in the same approximation reads
\begin{align} \label{eq:HIEM}
   \hat{H}^{EM}_I =  \frac{1}{2} n g^{EM}  \left(  N +  \sum_{\bm{k} \neq \bm{0}}  \hat{A}_{\bm{k}} \right).
\end{align}
The difference between the long-range nature of the quantum gravitational case \eqref{eq:HIQG} and the short-range nature of the (also quantum) electromagnetic case \eqref{eq:HIEM} is manifested in the second term (excitations) of the above expressions: in the gravity case we have a $k$ dependent coupling while in the electromagnetic case we have a constant coupling.

Finally, in sharp contrast to the quantum gravity case, for classical gravity we find that the chemical potential is updated to $\mu \rightarrow \mu - n g^G_0$ and the interaction Hamiltonian \eqref{eq:HICG} can be written as just
\begin{align} 
\label{eq:HCG}
        \hat{H}_I^{CG} =  n_T N_T g^G_0,
\end{align}
where $N_T := \sum_{\bm{k}\neq 0} \langle \hat{a}^{\dagger}_{\bm{k}} \hat{a}_{\bm{k}}\rangle $ is the number of non-condensed atoms and $n_T := N_T / \mathcal{V}$ is their density. That is, the interaction Hamiltonian  is simply a constant term. The radically different behaviour of the interaction between the classical and quantum gravity cases we consider is what allows us to discern the two cases. In particular, as we will see in a moment, the difference can be read from the energy spectrum and heat capacity. 

\emph{Bogolyubov transformation ---} We next apply a linear symplectic transformation on phase space to diagonalize the canonical Hamiltonians:
    $\hat{b}_{\bm{k}} = u_{\bm{k}} \hat{a}_{\bm{k}} + v^{\ast}_{-\bm{k}} \hat{a}^{\dagger}_{-\bm{k}}$, and
    $\hat{b}^{\dagger}_{\bm{k}} = u^{\ast}_{\bm{k}} \hat{a}^{\dagger}_{\bm{k}} + v_{-\bm{k}} \hat{a}_{-\bm{k}}$, where $|u_{\bm{k}}|^2 - |v_{-\bm{k}}|^2 = 1$. For the quantum gravity case \eqref{eq:HIQG} , this results in the grand-canonical Hamiltonian
\begin{align} \label{eq:KQG}
    \hat{K}^{QG} = E_{0}^{QG} + \sum_{\bm{k}\neq 0} \epsilon_{\bm{k}}^{QG} \hat{b}^{\dagger}_{\bm{k}} \hat{b}_{\bm{k}},
\end{align}
with excitation energies
\begin{align} \label{eq:EigenEnerQ}
    \epsilon_{\bm{k}}^{QG} = \sqrt{ \bigg( \frac{\hbar^2 k^2}{2 m} - \mu + n \big( 2 g^{EM} + g^{0}_k + g^{G}_k \big) \bigg)^2 - \bigg( n ( g^{EM} +  g^{G}_k) \bigg)^2} \nonumber \\
\end{align}
and ground state energy
\begin{equation} \label{eq:E0QG}
    E_0^{QG}  = E_0  + \frac{1}{2} n N g^{G}_0   + \frac{1}{2} \sum_{k \neq 0} \Bigg[ \epsilon^{G}_k -  n g^{G}_k   \Bigg],
\end{equation}
with
\begin{align}
    E_0  =  \frac{1}{2} n N  g^{EM}  +  \frac{1}{2}  \, \sum_{k \neq 0} \left[\mu - \frac{\hbar^2 k^2}{2 m}   - 2 n g^{EM} - n g^G_0 \right].
\end{align}
For the classical gravity case  \eqref{eq:HICG}, we find
\begin{align} \label{eq:KCG}
    \hat{K}^{CG} = E_{0}^{CG} + \sum_{\bm{k}\neq 0} \epsilon_{\bm{k}}^{CG} \hat{b}^{\dagger}_{\bm{k}} \hat{b}_{\bm{k}},
\end{align}
where
\begin{align}
     \epsilon_{k}^{CG} = \sqrt{\bigg( \dfrac{\hbar^2 k^2}{2 m} - \mu + 2 n g^{EM} + n g^G_0 \bigg)^2 - \left(n g^{EM}\right)^2 },
\end{align}
and
\begin{equation} \label{eq:E0CG}
    E_0^{CG}  =   E_0 + n_T  N_T g^G_0 + \frac{1}{2}  \, \sum_{k \neq 0} \epsilon^{CG}_k.
\end{equation}

\emph{Energy spectrum ---} In order to obtain the final form of the energy spectrum, we  apply the condition that the excitations must be gapless:
\begin{equation} \label{eq:dispquant}
        \lim_{\mathbf{k} \rightarrow 0 } \epsilon^{QG}_{\bm{k}} = 0 ~~\text{
  and}~~ \lim_{\mathbf{k} \rightarrow 0 } \epsilon^{CG}_{\bm{k}} = 0.
\end{equation}
 This is typically considered a necessary step in order to have a BEC according to the Nambu-Goldstone theorem \cite{1960PhRvL...4..380N,1961NCim...19..154G}, and allows one to infer the chemical potential $\mu$ \cite{2006PhRvA..73f3612Y}.
 Assuming that the electromagnetic interactions are repulsive $g^{EM}>0$, for the classical gravity interaction Hamiltonian \eqref{eq:HICG} we obtain $\mu = n (g^{EM} + g^G_0)$, 
 leaving us with
\begin{align} \label{eq:ekCG1}
\epsilon_{\bm{k}}^{CG} = \hbar k \sqrt{ \frac{\hbar^2 k^2}{ 4 m^2} + \frac{ n  g^{EM}}{m} }.
\end{align}
This result is just the standard excitation energies for a Bose gas with  electromagnetic interactions in the approximation of the  Bogolyubov method \cite{PitaevskiiBook}.

For the quantum gravity interaction Hamiltonian \eqref{eq:HIQG} and assuming again $g^{EM}>0$,  the situation is more subtle. We will discuss separately the case when the electromagnetic coupling is larger than the gravitational coupling and vice versa. With  $g^{EM} > |g^{G}_0|$, we find  $\mu = n ( g^{EM} + g^{G}_0)$ and obtain
\begin{align} \label{eq:ekQG1}
\epsilon_{\bm{k}}^{ QG} = \hbar k \sqrt{ \frac{\hbar^2 k^2}{ 4 m^2} + \frac{ n  g^{EM} }{m} + \frac{ n    g^{G}_k}{m}}.
\end{align}
Whereas, if $g^{EM} < |g^{G}_0|$ then $\mu = 3 n ( g^{EM} + g^{G}_0)$, in which case we obtain
 \begin{align}\label{eq:ekQG2}
\epsilon_{\bm{k}}^{QG} =  \sqrt{\left( \frac{\hbar^2 k^2}{2m} - 2 n (g^{EM} + g^{G}_0)\right) \left( \frac{\hbar^2 k^2}{2m} - 2 n (g^{G}_0 - g^{G}_k)\right) }. \nonumber \\
\end{align}
Note that the expression under the root never becomes negative. This is because $g^{G}$ is always negative and $|g_0^{G}| > g^{EM}$ and $|g_0^{G}| > |g^{G}_k|$. 
\\
\\
Some comments are in order. In \eqref{eq:ekQG1}, there will be phononic-like excitations when $\hbar k \ll   m c_s$  and $n_0 g^{G}_k / m \ll c_s^2$, where we define $c^2_s = n_0 g^{EM} / m$. In contrast, in \eqref{eq:ekQG2}, it is never possible to obtain phononic excitations given the form of $g^{G}_k$ in \eqref{eq:Vq}. This means that if gravity becomes greater than the electromagnetic interactions, instead of phonons (type-A Nambu-Goldstone bosons)\footnote{See Supplementary Material for more detail.}, one should see the emergence of type-B Nambu-Goldstone bosons sourced by the quadratic dispersion relation. Additionally, the fact that  quantum gravity produces Type-B NGBs also presents a clear-cut criterion to tell it apart from EM-sourced-dispersion, as the latter only ever produces Type-A NGBs (phonons). 

\medskip

\emph{Heat capacity --- } 
The heat capacity can be evaluated as $    c_V = \partial E / \partial T$,
where $E = \langle \hat{K}^G \rangle + \mu N$ is the internal energy, with $\hat{K}^G$ being $\hat{K}^{QG}$ or $\hat{K}^{CG}$. Given that $\langle b^{\dagger}_k b_k \rangle = \int dk \, 1/(\exp(\beta \epsilon_k) - 1)$, the specific heat capacity $c_V$ is then
\begin{equation} \label{eq:cv}
    c_V = \frac{1}{k_B T^2} \sum_{k \neq 0} \dfrac{\epsilon_k^2 e^{\beta \epsilon_{k} } \, }{ \, (e^{\beta \epsilon_{k} } -1)^2},
\end{equation}
where $\beta := 1 / (k_B T)$ and $\epsilon_k$ is $\epsilon^{QG}_k$ 
 for quantum gravity and $\epsilon^{CG}_k$ for classical gravity, assuming sufficiently low temperatures. Since the form of \eqref{eq:cv} is the same for both quantum and classical gravity, any  difference comes entirely from the respective $\epsilon_k$ calculated above. In fact, since $\epsilon_k^{CG}$ is the same as that for a non-self gravitating BEC with electromagnetic interactions, the heat capacity of a BEC is  unaffected by a classical gravity interaction of the form \eqref{eq:HICG}. Therefore, we see that the heat capacity, a macroscopic thermodynamical observable, will in principle cleanly distinguish between the two cases we consider: In the classical gravity case, the heat capacity will be unaffected and will be the same as if we had neglected the self-gravitational interaction, which is not the case for the quantum gravitational self-interaction. This is due to  the spacing of the energy eigenvalues being only affected by quantum gravity: they follow  \eqref{eq:ekQG1} or \eqref{eq:ekQG2}
 rather than  \eqref{eq:ekCG1}.

\emph{Size of the effect --- } 
We now consider what sort of experimental parameters could be  required to evidence quantum gravity through measurements of the  heat capacity. Since any differences come through the energy eigenvalues, as discussed above, we first discuss how measurements of just these eigenvalues could also be used to distinguish quantum and classical gravity. For example, we could consider a $0.1\%$ deviation from  the standard values $\epsilon^{CG}_{\bm{k}}$ as the minimum threshold  for  a measurable result and evidence of quantum gravity. That is, we can consider what experimental parameters are required for $ \Delta \epsilon_{\bm{k}} / \epsilon^{CG}_{\bm{k}} \times 100\%  \geq 0.1\%$, where $\Delta \epsilon_{\bm{k}} := |\epsilon^{QG}_{\bm{k}} - \epsilon^{CG}_{\bm{k}}|$. For now, we assume that the electromagnetic  interactions can be set negligible compared to quantum gravity using Feshbach resonances, and return to this later. Then, from \eqref{eq:ekQG2}, if we ignore $g^G_k$ for simplicity, the  condition for the energy eigenvalues to be distinguishable is satisfied when
\begin{align}
100\, n \, g_0^G &\gtrapprox 0.1 \hbar^2 k^2 / ( 4m^2)\\ \label{eq:NL}
\implies N L &\gtrapprox  \pi^2 \hbar^2 |\bm{n}_k|^2 / (2380 G m^3).
\end{align}
Taking a $^{174}{\mathrm{Yb}}$ Bose gas, this sets the approximate condition $N L \gtrapprox 2.9 \times 10^{14}\, |\bm{n}_k|^2 \,\mathrm{m^{-3}}$. Clearly, it is preferable  to measure the $|\bm{n}_k| = 1$ mode. Then, an approximate $0.1\%$ deviation would require, for example, $N \approx 10^{15}$ and $L \approx 1 \,\mathrm{cm}$. This is above what has been achieved in experiments - with $N=10^9$ Hydrogen atoms being the greatest \cite{PhysRevLett.81.3811}, with the condensate being $0.5\,\mathrm{cm}$ in length and $15\,\mathrm{\mu m}$ in diameter; and $10^8$ Sodium atoms \cite{doi:10.1063/1.2424439} in a BEC with axial and radial sizes of $1\,\mathrm{mm}$ and $20\,\mathrm{\mu m}$.   It also requires a relatively high density compared to typical experiments, resulting in a theoretical half-life of around $1\,\mathrm{ms}$ due to three-body decay, where the rate of change in density is $d n(t) / dt = - L n^3(t)$, with $L$ the rate coefficient, which is of order $10^{-41}\,\mathrm{m^6 \,s^{-1}}$ for $^{174}{\mathrm{Yb}}$ \cite{PhysRevA.79.021601}. 

Now we consider what experimental parameters might be required to distinguish quantum and classical gravity in measurements of the \emph{thermodynamic}  quantity $c_V$. We again consider a $0.1\%$ deviation in these values as evidence of quantum gravity, i.e.\ $100\% \times \Delta c_V / c_V^{CG} \geq 0.1\%$, where $\Delta c_V := |c_V^{QG} - c_V^{CG}|$, with $c_V^{QG}$ and $c_V^{CG}$ defined in \eqref{eq:cv} for the respective energy eigenvalues. In this case, similar but slightly more demanding parameters are found to be required than in the measurement of the lowest energy mode, with for example $N \approx 10^{16}$, $L \approx 1 \,\mathrm{cm}$ and a temperature $T \approx 0.01\,\mathrm{pK}$ providing a $0.1\%$ deviation from a standard  (classical gravity) heat capacity of  $ c_V^{CG} / k_B \approx 3.164$\footnote{Despite the high mass, the non-relativistic, perturbative treatment used to derive the heat capacity is still valid, as demonstrated in the Supplementary Material.}. Again this is very much pushing current experiments, with a temperature of order $10\,\mathrm{pK}$ the smallest  observed so far in a Bose gas \cite{PhysRevLett.127.100401}. However, although the required numbers appear slightly more demanding, it may be experimentally preferable to measure the heat capacity, a macroscopic thermodynamic quantity,  rather than attempting to single out only the  first quasiparticle mode and measure its energy.

We now discuss  the  electromagnetic self--interaction contributions to the above measurements.  Being long-range, the gravitational interaction scales very differently to the electromagnetic one: since the latter is assumed to be a contact potential, $g^{EM}$ is independent of $L$, whereas $g^{G}$ scales with $L^2$ from \eqref{eq:Vq}. This difference makes it possible to distinguish the two cases by measuring the functional dependence of the heat capacity on $L$. Specifically,   a spatially large condensate is preferable (with the number of atoms  making no difference). Taking a $^{174}{\mathrm{Yb}}$ BEC with $L \approx 1 \,\mathrm{cm}$, results in $g^{EM}/g^{G}_0 \approx 10^{11} $. However,  another aspect of BECs can be utilised:  Feshbach resonances such that an appropriate magnetic (or optical in the case of Strontium) field applied to a BEC can  reduce the strength of the electromagnetic interactions. The challenge would be to use a precise enough magnetic field such that fluctuations in its field strength are minimal  \cite{Howl:2020isj}.

\emph{Conclusions --- } We have studied how the gravitational self-interaction of a uniform BEC affects its  heat capacity in the approximation of the Bogolyubov method and non-relativistic gravity. We find that this thermodynamic quantity is unaffected when the gravitational interaction is classical in the spirit of Schr\"{o}dinger-Newton theory - see \eqref{eq:HICG}. In contrast, if the gravitational interaction is quantized, a gravitational term that is  momentum-dependent enters the excitation energies -- see \eqref{eq:ekQG1} and \eqref{eq:ekQG2} -- which also modifies the  heat capacity. The two cases of classical and quantum gravity can thus be distinguished in principle by measuring these modifications. This is our main result.

In practice, observing these modifications would require substantial improvements in the state-of-the-art of current BEC experiments. For example, while further studies are needed  to assess the full feasibility of our idea, our estimates suggest that a gain of at least $\sim 7 $ orders of the number of atoms in the condensate are required. Possible approaches to significantly increase state-of-the-art BEC atom numbers include using multi-frequency traps to trap more atoms in a Magneto-Optical Trap (MOT) \cite{sinclair1994improved,Cao_2012,howl2023gravitationallyinducedentanglementcoldatoms} and then  optical barriers to cool the atoms to a BEC, avoiding evaporative cooling, which typically results in orders of magnitude losses in atom numbers from MOTs \cite{Raizen2005,raizen2014magneto,Hillberry_2023}. Furthermore, operating BEC experiments in space can lead to increased atom numbers \cite{aveline2020observation,becker2018space,howl2023gravitationallyinducedentanglementcoldatoms}. We also note that  proposals based on measurements of quantum-information quantities, such as entanglement, also require substantial  gains compared to the state-of-the-art. For example, the Bose et al.\ proposal \cite{PhysRevLett.119.240401}, requires Schr\"{o}dinger cat-like states of matter that are around $ 8$ orders greater in mass than the current record for such states \cite{fein2019quantum}, and gravitationally-induced entanglement with  cold atom interferometers requires similar levels of atom numbers when the  interferometers are not quantum squeezed \cite{howl2023gravitationallyinducedentanglementcoldatoms}. In contrast to such proposals, the proposal considered here does not necessitate coherent control or the application of quantum operations and measurements.

Since high numbers of atoms are needed to detect the quantum gravity signal, it is possible that denser macroscopic quantum systems, such as liquid Helium, could be more suited to our proposed test. This is further supported by the straightforward application of our derived condition  
$ N L \approx 10^{14}$ for quantum gravity to have a significant effect in a BEC -- when applied to liquid helium, this condition is easily satisfied \cite{Sato_2012}, highlighting its potential as a promising platform for such investigations. 
 Extending our analysis to a superfluid may be non-trivial since the microscopic description of a superfluid is still being debated \cite{nozieres2018theory} and there are no known  Feshbach resonances.   We note that neutron stars have also been proposed as systems that may undergo Bose-Einstein condensation \cite{haskell2018superfluidity}, raising the intriguing possibility that gravitational wave or electromagnetic signals emitted by such astronomical objects could provide insights into properties like heat capacity and, in turn, potentially offer  evidence of quantum gravity.

The approach explored here focuses on providing evidence for quantum gravity through measurements of macroscopic, thermodynamical observables. Interestingly, there is a conceptual link to methods that rely on quantum-information principles for demonstrating the quantum nature of gravity: In the Bogolyubov approximation, the quantum gravity Hamiltonian \eqref{eq:HIQG} contains off diagonal terms $\hat{a}_{\bm{k}} \hat{a}_{-\bm{k}} + h.c.$, which will generate entanglement in the atomic modes, whereas no such terms exist in the classical gravity Hamiltonian \eqref{eq:HCG}. This suggests a connection to previous considerations of  how heat capacity can be used to evidence entanglement \cite{HeatCapacityEnt}. Investigating further these connections may reveal whether measuring the change in the  heat capacity could rule out  forms of classical gravity beyond semi-classical gravity, such as those based on stochasticity \cite{CGReview}.

\subsubsection*{Acknowledgements}

We acknowledge support from the ID\# 61466 and ID\# 62312 grants from the John Templeton Foundation, as part of the ``Quantum Information Structure of Spacetime (QISS)'' project (\hyperlink{http://www.qiss.fr}{qiss.fr}). This research was funded in whole or in part by the Austrian Science Fund (FWF)[10.55776/F71] and
[10.55776/COE1]. For open access purposes, the author has applied a CC BY public copyright license to any author accepted manuscript version arising from this submission. RH also acknowledges the John Templeton Foundation grant ID\# 62420.

\bibliography{references}

\clearpage
\newpage

\onecolumngrid

\appendix

\pagestyle{empty}

\section*{Supplementary Material}

\subsection{Fourier series of gravitational potential of a cube} \label{app:PotentialCube}

Here, we analyse the Fourier series of the gravitational potential of a boxed Bose gas. Specifically, we consider the approximation \eqref{eq:Vq} from the expression \eqref{eq:VkCube}:
\begin{align} \label{eq:VkCubeapp}
    V^{G}_{\bm{k}} 
    &= \int^{\frac{L}{2}}_{-\frac{L}{2}} dx \int^{\frac{L}{2}}_{-\frac{L}{2}} dy \int^{\frac{L}{2}}_{-\frac{L}{2}} dz   \frac{- G m^2 e^{-i  (k_x x + k_y y + k_z z)}  }{\sqrt{x^2 + y^2 +z^2} },
\end{align}
where $k_x := 2 \pi n_x /L$, $k_y := 2 \pi n_y /L$ and $k_z := 2 \pi n_z /L$.

First, following \cite{CubeCoulomb} for deriving the Coulomb potential of a cube, we use the identity:
\begin{align}
    \int^{\infty}_0 dt e^{-a t^2 } &=\frac{\sqrt{\pi}}{2} \frac{1}{\sqrt{a}},\\
    \implies \frac{1}{\sqrt{x^2 + y^2 + z^2}} &= \frac{2}{\sqrt{\pi}} \int^{\infty}_0 dt e^{-(x^2 + y^2 + z^2) t^2 } 
\end{align}
We can then write:
\begin{align}
    V^G_{\bm{k}} &= \frac{-2G m^2}{\sqrt{\pi} } \int^{\infty}_0 dt \int^{L/2}_{-L/2} dx \int^{L/2}_{-L/2} dy \int^{L/2}_{-L/2} dz    e^{ -i  (k_x x + k_y y + k_z z)} e^{- (x^2 + y^2 + z^2) t^2 } \\
    &= \frac{-2 G m^2}{\sqrt{\pi}} \int^{\infty}_0 dt \int^{L/2}_{-L/2} dx e^{ -i  k_x x - x^2 t^2}  \int^{L/2}_{-L/2} dy e^{ -i  k_y y - y^2 t^2} \int^{L/2}_{-L/2} dz  e^{ -i  k_z z - z^2 t^2} 
\end{align}
Integrating over $x$, $y$ and $z$, we obtain:
\begin{align} \label{eq:fullVGapp}
    V^G_{\bm{k}} &= -2 \pi G m^2\int^{\infty}_0 dt \frac{e^{- k^2 / (4 t^2)}}{t^3} \mathrm{Re}\,[\erf(L t/2 + i k_x / (2 t))] \,\mathrm{Re}[\erf(L t/2 + i k_y / (2 t))] \, \mathrm{Re}[\erf(L t/2 + i k_z / (2 t))],
\end{align}
where $k^2 := k_x^2 + k_y^2 + k_z^2$.  We now change variables to $v = 1/ (L t)$  so that:
\begin{align} \label{eq:FinalInt}
    V^G_{\bm{k}} &= -2 \pi G m^2 L^2  \int^{\infty}_0 dv\, v \, e^{- \pi^2 v^2 n^2_k } \mathrm{Re}[\erf(1 /(2 v) + i \pi n_x v  )] \mathrm{Re}[\erf(1 /(2 v) + i \pi n_y v  )] \mathrm{Re}[\erf(1 /(2 v) + i 
 \pi n_z v  )]\\
    &:=-2 \pi G m^2 L^2 \int^{\infty}_0 dv f(v)
\end{align}
where $n^2_k := n_x^2 + n_y^2 + n_z^2$. Since $n_x, n_y $ and $n_z$ are integers, for all $n_x$, $n_y$ and $n_z \neq 0$, $f(v)$ is completely dominated by the term $v \, e^{- \pi^2 v^2 n^2_k }$. We can thus approximate $f(v)$ by just $f(v) \approx v\, e^{- \pi^2 v^2 n^2_k }$.\footnote{A slightly better approximation can be obtained by expanding $\mathrm{Re}[\erf(1 /(2 v) + i \pi n_x v  )] \mathrm{Re}[\erf(1 /(2 v) + i \pi n_y v  )] \mathrm{Re}[\erf(1 /(2 v) + i  \pi n_z v  )]$ to first order in $v$.} Integrating this approximation, we obtain
\begin{align}
     V^G_{\bm{k} \neq 0} \approx -2 \pi G m^2 L^2  \frac{1}{2 \pi^2 n^2_k} \equiv - \frac{4 \pi G m^2}{k^2},
\end{align}
which is what we would obtain for a sphere with volume $L^3$. Numerically, we find that this solution is indeed a very good approximation to \eqref{eq:VkCubeapp} when $\bm{k} \neq 0$. 

All that remains  is to find \eqref{eq:VkCubeapp} when $\bm{k} = 0$. Setting $\bm{k} = 0$ in \eqref{eq:fullVGapp}, we need to calculate:
\begin{align} \label{eq:VG0app}
    V^G_{0} &= -2 \pi G m^2 \int^{\infty}_0 dt \frac{1}{t^3} \erf^3(L t/2 ) \\
    &=  \frac{- \pi G m^2 L^2}{2 } \int^{\infty}_0 dw \frac{1}{w^3} \erf^3(w ).
\end{align}
Integrating by parts twice (first with $u = \erf^3 w$ and second with $u = \erf^2 w $), and using the identity:
\begin{align}
    \int^{\infty}_0 e^{-p x^2} \erf(ax) \erf(b x) = \frac{1}{\pi p} \arctan\left(\frac{a b}{\sqrt{p (a^2 + b^2 + p)}}\right),
\end{align}
we find (see also \cite{CubeCoulomb}):
\begin{align}
    V^G_0 = - \frac{\pi G m^2 L^2}{2}  \left[ \frac{12}{\pi} \mathrm{arcsinh}\left(\frac{1}{\sqrt{2}}\right) - 1\right] \approx - 2.38 G m^2 L^2. 
\end{align}

\subsection{Bogolyubov method for quantum gravity} \label{app:KQGDerivation}

Here we detail how the diagonalized grand canonical Hamiltonian for quantum gravity \eqref{eq:KQG} is derived from the original non-diagonalized version \eqref{eq:K} in the Bogolybov method.

We start with $\hat{H}_I^{QG}$ as given in \eqref{eq:HIG}: 
\begin{align} \label{eq:HIQGapp}
    \hat{H}_{I}^{QG} = \frac{1}{2}\, \!\! \int \!\!\! d^3 \mathbf{r'} \!\! \int \!\!\! d^3 \mathbf{r} \;  \hat{\Psi}^{\dagger} (\mathbf{r}) \hat{\Psi}^{\dagger} (\mathbf{r'}) V_I(\bm{r} - \bm{r'}) \hat{\Psi} (\mathbf{r'}) \hat{\Psi} (\mathbf{r}),
\end{align}
where $V_I(\bm{r} - \bm{r'}) = -  \frac{G m^2} { |\bm{r-r'}|}$.

As in the main text, we assume a Bose gas in a box of sides $L$ and with periodic boundary conditions. The field operator $\hat{\Psi}(\mathbf{r})$ can then be written as:
\begin{align} 
    \hat{\Psi} (\mathbf{r}) &= \frac{1}{\sqrt{\mathcal{V}}} \sum_{\bm{k}} e^{i \bm{k}\cdot\bm{r}} \hat{a}_{\bm{k}}\\ \label{eq:Psiapp}
    &\equiv \frac{1}{\sqrt{\mathcal{V}}} \hat{a}_0 + \frac{1}{\sqrt{\mathcal{V}}} \sum_{\bm{k} \neq 0} e^{i \bm{k}\cdot\bm{r}} \hat{a}_{\bm{k}},
\end{align}
where  $\bm{k} = 2 \pi \bm{n}_k / L$,  $\bm{n}_k = (n_x, n_y, n_z)$, and $n_x,~n_y,~n_z$ integers ranging from $-\infty$ to $+\infty$. 

We also likewise expand the potential $V_I(\bm{r} - \bm{r'})$ as:
\begin{align} \label{eq:VIapp}
    V_I(\bm{r} - \bm{r'}) = \frac{1}{\mathcal{V}}\sum_{\bm{k}} V^G_{\bm{k}} \, e^{i\bm{k}\cdot (\bm{r} - \bm{r'}) }.
\end{align}
Plugging \eqref{eq:VIapp} and \eqref{eq:Psiapp} into \eqref{eq:HIQGapp}, we would obtain an expansion of $\hat{H}^{QG}_I$ with terms consisting of no $\hat{a}_0$ operators, one $\hat{a}_0$ operator, two $\hat{a}_0$ operators, three  $\hat{a}_0$ operators and four. Since we will later apply the Bologyubov approximation, where we replace $\hat{a}_0$ with $\sqrt{N_0}$ and take $\sqrt{N_0} \gg 1$, we ignore terms with zero and one $\hat{a}_0$ operator. The terms with three $\hat{a}_0$ operators can also be shown to vanish  : Integrating over $\bm{r}$ and $\bm{r'}$ such terms would yield a Kronecker delta that sets $\bm{k} = 0$, which is not possible 
 - see, for example, \cite{D_A_W_Hutchinson_2000} for more detail.

This leaves us with:
\begin{align}
    \hat{H}_I^{QG} = \frac{1}{\mathcal{V}^3} \sum_{\bm{q}} V^G_{\bm{q}} \frac{1}{2} \int d^3 \bm{r} \int d^3 \bm{r'} e^{i \bm{q}\cdot(\bm{r} - \bm{r'})} \Bigg[ &\hat{a}_0^{\dagger} \hat{a}_0^{\dagger} \hat{a}_0 \hat{a}_0 + \sum_{\bm{k},\bm{k'}} \bigg[ \hat{a}_0 \hat{a}_0 e^{-i (\bm{k}\cdot\bm{r'} + \bm{k'}\cdot\bm{r})} \hat{a}_{\bm{k}}^{\dagger}  \hat{a}_{\bm{k'}}^{\dagger} + \hat{a}^{\dagger} _0 \hat{a}^{\dagger} _0 e^{i (\bm{k}\cdot\bm{r'} + \bm{k'}\cdot\bm{r})} \hat{a}_{\bm{k}} \hat{a}_{\bm{k'}}\\
    &+ \hat{a}_0^{\dagger} \hat{a}_0 \left(e^{i(\bm{k'} - \bm{k})\cdot\bm{r}}  + e^{i(\bm{k'} - \bm{k}).\bm{r'}}  + e^{i(\bm{k'}\cdot\bm{r'}-\bm{k}\cdot\bm{r})}   + e^{i(\bm{k'}\cdot\bm{r} - \bm{k}\cdot\bm{r'})}  \right)  \hat{a}^{\dagger}_{\bm{k}} \hat{a}_{\bm{k'}}\bigg] \Bigg].
\end{align}
Integrating over $\bm{r}$ and $\bm{r'}$ picks up Kronecker deltas, for example:

\begin{align}
    \int^{\frac{L}{2}}_{-\frac{L}{2}} \int^{\frac{L}{2}}_{-\frac{L}{2}} \int^{\frac{L}{2}}_{-\frac{L}{2}} d^3 \bm{r}\,  e^{i \bm{r}.(\bm{q} + \bm{k} - \bm{k'})} &= \int^{\frac{L}{2}}_{-\frac{L}{2}} \int^{\frac{L}{2}}_{-\frac{L}{2}} \int^{\frac{L}{2}}_{-\frac{L}{2}} \,d x \,dy \, d z  \, e^{i x (q_x + k_x - k'_x)} e^{i y (q_y + k_y - k'_y)} e^{i z( q_z + k_z - k'_z)} = \mathcal{V} \delta_{\bm{q},\bm{k'}-\bm{k}}.
\end{align}
 We thus obtain
\begin{align}
    \hat{H}_I^{QG} = \frac{1}{2 \mathcal{V}} \Bigg[ V^G_0\, \hat{a}_0^{\dagger} \hat{a}_0^{\dagger} \hat{a}_0 \hat{a}_0 + \sum_{\bm{k}} \bigg[ V^G_{\bm{k}} \Big(
  \hat{a}_0 \hat{a}_0 \hat{a}_{\bm{k}}^{\dagger}  \hat{a}_{-\bm{k}}^{\dagger} + \hat{a}^{\dagger}_0 \hat{a}^{\dagger} _0  \hat{a}_{\bm{k}} \hat{a}_{-\bm{k}}\Big)
    + \hat{a}_0^{\dagger} \hat{a}_0 \Big(2 V^G_0 +  V^G_{\bm{k}}  + V^G_{-\bm{k}}\Big) \hat{a}_{\bm{k}}^{\dagger} \hat{a}_{\bm{k}}   \bigg]\Bigg].
\end{align}
We now apply the Bogolyubov approximation and replace $\hat{a}_0$ with $\sqrt{N}$, such that@
\begin{align}
    \hat{H}_I^{QG} = \frac{V^G_0 N^2}{2 \mathcal{V}}    &+  \frac{N}{2 \mathcal{V}} \sum_{\bm{k} \neq \bm{0}}    \bigg(  2(V^G_{0} + V^G_{\bm{k}})   \hat{a}^{\dagger}_{\bm{k}} \hat{a}_{\bm{k}} +  V^G_{\bm{k}} \left(\hat{a}^{\dagger}_{\bm{k}} \hat{a}^{\dagger}_{-\bm{k}}  +  \hat{a}_{\bm{k}} \hat{a}_{-\bm{k}}\right) \bigg),
\end{align}
where, from \eqref{eq:VkCube}, we have used $V^G_{-\bm{k}} = V^G_{\bm{k}}$. Replacing $V^G_k$ with $g^{QG}_k$, the above results in \eqref{eq:HIQG}.

The above method can also be applied to the grand canonical Hamiltonian \eqref{eq:K} containing the kinetic term and electromagnetic interactions \cite{PitaevskiiBook}, resulting in:
\begin{align}
    \hat{K}^{QG} = \frac{1}{2} \left(g^{EM} + g^{QG}_0\right) n N &+ \sum_{\bm{k}} \left(\frac{\hbar^2 k^2}{2m} - \mu\right) \hat{a}^{\dagger}_{\bm{k}} \hat{a}_{\bm{k}} \\&+ \frac{1}{2} n \sum_{\bm{k}\neq 0}   \bigg( \left(4 g^{EM} +  2( g^{QG}_0 + g^{QG}_k) \right) \hat{a}^{\dagger}_{\bm{k}} \hat{a}_{\bm{k}} + \left(g^{EM}  + g^{QG}_k\right) \left(\hat{a}^{\dagger}_{\bm{k}} \hat{a}^{\dagger}_{-\bm{k}}  +  \hat{a}_{\bm{k}} \hat{a}_{-\bm{k}}\right) \bigg),
\end{align}
where, for simplicity, we have ignored the electromagnetic term $ m \left(g^{EM} n \right)^2 / (2k^2)$ that is often implemented due to going beyond the Born approximation since this term only contributes to the ground state energy, which  does not contribute to the internal energy or heat capacity \cite{PitaevskiiBook}.

We now diagonalize $\hat{K}^{QG}$ with the Bogolyubov transformation:
\begin{align}
    \hat{a}_{\bm{k}} &= u_{\bm{k}} \hat{b}_{\bm{k}} - v^{\ast}_{-\bm{k}} \hat{b}^{\dagger}_{-\bm{k}},\\
    \hat{a}^{\dagger}_{\bm{k}} &= u^{\ast}_{\bm{k}} \hat{b}^{\dagger}_{\bm{k}} - v_{-\bm{k}} \hat{b}_{-\bm{k}},
\end{align}
 where:
 \begin{align} \label{eq:uk}
     u_{\bm{k}} &:= \sqrt{\frac{\hbar^2 k^2 / (2m) - \mu +  n\left( 2 g^{EM} + g^{QG}_k + g^{QG}_0 \right)}{2\epsilon_k^{QG}} + \frac{1}{2} },\\ \label{eq:vk}
     v_{\bm{k}} &:= \sqrt{\frac{\hbar^2 k^2 / (2m) - \mu +  n\left( 2 g^{EM} + g^{QG}_k + g^{QG}_0 \right)}{2\epsilon_k^{QG}} - \frac{1}{2}},\\ \label{eq:ekQGapp}
      \epsilon_{\bm{k}}^{QG} &= \sqrt{ \bigg( \frac{\hbar^2 k^2}{2 m} - \mu +  n \big(  2 g^{EM} + g^{QG}_0 + g^{QG}_k \big) \bigg)^2 - \bigg( n ( g^{EM} +  g^{QG}_k) \bigg)^2},
 \end{align}
to end up with \eqref{eq:KQG}:
\begin{align} \label{eq:KQGapp}
    \hat{K}^{QG} = E_{0}^{QG} + \sum_{\bm{k}\neq 0} \epsilon_{\bm{k}}^{QG} \hat{b}^{\dagger}_{\bm{k}} \hat{b}_{\bm{k}},
\end{align}
where $E_0^{QG}$ is defined in the main text - see \eqref{eq:E0QG}.

\subsection{ Bogolyubov method for classical gravity} \label{app:KCGDerivation}

Here we detail how the diagonalized grand canonical Hamiltonian for classical gravity \eqref{eq:KCG} is derived from the original non-diagonalized version \eqref{eq:K} in the Bogolybov method.

We start with $\hat{H}_I^{CG}$ as given in \eqref{eq:HICG}: 
\begin{align} \label{eq:HICGapp}
    \hat{H}^{CG}_{I} =  \int \!\! d^3 \mathbf{r'} \int \!\! d^3 \mathbf{r}  \, V_I(\bm{r} - \bm{r'}) \langle  \hat{\Psi}^{\dagger} (\mathbf{r'}) \hat{\Psi}(\mathbf{r'}) \rangle  \, \hat{\Psi}^{\dagger} (\mathbf{r}) \hat{\Psi} (\mathbf{r}).
\end{align}
where $V_I(\bm{r} - \bm{r'}) = -  \frac{G m^2} { |\bm{r-r'}|}$.

As in with quantum gravity, we plug in 
\begin{align} 
    \hat{\Psi} (\mathbf{r}) &= \frac{1}{\sqrt{\mathcal{V}}} \hat{a}_0 + \frac{1}{\sqrt{\mathcal{V}}} \sum_{\bm{k} \neq 0} e^{i \bm{k}.\bm{r}} \hat{a}_{\bm{k}},\\
    V_I(\bm{r} - \bm{r'}) &= \frac{1}{\mathcal{V}} \sum_{\bm{k}} V_{\bm{k}} \, e^{i\bm{k}\cdot (\bm{r} - \bm{r'}) },
\end{align}
such that:
\begin{align}
    \hat{H}^{CG}_{I} =  \frac{1}{\mathcal{V}^3} \sum_{\bm{q}} V_{\bm{q}} \int \!\! d^3 \mathbf{r'} \int \!\! d^3 \mathbf{r}  \, e^{i \bm{q}\cdot(\bm{r} - \bm{r'})} \Bigg[\bigg(\big\langle  \hat{a}_0^{\dagger} \hat{a}_0 \big\rangle+ \sum_{\bm{k}, \bm{k'} \neq 0} e^{i ( \bm{k}- \bm{k'})\cdot\bm{r'}} \big\langle \hat{a}^{\dagger}_{\bm{k'}} \hat{a}_{\bm{k}} \big\rangle\bigg) \bigg( \hat{a}^{\dagger}_0 + \sum_{\bm{k'}\neq 0} e^{-i\bm{k'}\cdot\bm{r} } \hat{a}^{\dagger}_{\bm{k'}} \bigg) \bigg( \hat{a}_0 + \sum_{\bm{k}\neq 0} e^{i\bm{k}\cdot\bm{r} } \hat{a}_{\bm{k}}\bigg) \Bigg],
\end{align}
where we have used $\langle \hat{\psi}(\bm{r})\rangle = 0$, with the average  taken over the quantum state of the entire Bose gas. 

As in the quantum gravity case, the terms with three $\hat{a}_0$ operators vanish since they do not conserve momentum and, since we will use the Bogolyubov approximation, we can ignore terms with no or just one $\hat{a}_0$ operator. This then leaves us with:
\begin{align} 
    \hat{H}^{CG}_{I} =  \frac{1}{\mathcal{V}^3} \sum_{\bm{q}} V_{\bm{q}} \int \!\! d^3 \mathbf{r'} \int \!\! d^3 \mathbf{r}  \, e^{i \bm{q}\cdot(\bm{r} - \bm{r'})} \Bigg[\big\langle  \hat{a}_0^{\dagger} \hat{a}_0 \big\rangle \hat{a}_0^{\dagger} \hat{a}_0 + \langle  \hat{a}_0^{\dagger} \hat{a}_0 \rangle\sum_{\bm{k},\bm{k'} \neq 0} e^{i (\bm{k} - \bm{k'})\cdot\bm{r}} \hat{a}^{\dagger}_{\bm{k'}} \hat{a}_{\bm{k}}+ \hat{a}_0^{\dagger} \hat{a}_0 \sum_{\bm{k}, \bm{k'} \neq 0} e^{i ( \bm{k}- \bm{k'})\cdot\bm{r'}} \langle \hat{a}^{\dagger}_{\bm{k'}} \hat{a}_{\bm{k}} \rangle \Bigg].
\end{align}
Integrating over $\bm{r}$ and $\bm{r}'$ then provides
\begin{align} 
    \hat{H}^{CG}_{I} =   \frac{V_0}{\mathcal{V}} \left(  \langle\hat{a}_0^{\dagger} \hat{a}_0 \big\rangle \hat{a}_0^{\dagger} \hat{a}_0 + \langle  \hat{a}_0^{\dagger} \hat{a}_0 \rangle  \sum_{\bm{k} \neq 0}  \hat{a}^{\dagger}_{\bm{k}} \hat{a}_{\bm{k}} + \hat{a}_0^{\dagger} \hat{a}_0 \sum_{\bm{k} \neq 0}  \langle \hat{a}^{\dagger}_{\bm{k}} \hat{a}_{\bm{k}} \rangle \right).
\end{align}
We now apply the Bogolyubov approximation and set $\hat{a}_0 \approx \sqrt{N}$. This then leaves us with:
\begin{align} 
    \hat{H}^{CG}_{I} &=
    \frac{V_0}{\mathcal{V}} \left(N^2 + N \sum_{\bm{k}\neq 0} \hat{a}^{\dagger}_{\bm{k}} \hat{a}_{\bm{k}} + N \sum_{\bm{k}\neq 0} \langle \hat{a}^{\dagger}_{\bm{k}} \hat{a}_{\bm{k}} \rangle \right)\\ \label{eq:HCGmu}
    &= \frac{V_0}{\mathcal{V}} N_T^2 + V_0 n \sum_{\bm{k}} \hat{a}^{\dagger}_{\bm{k}} \hat{a}_{\bm{k}},
\end{align}
with $N_T$ the total number of non-condensed  atoms. Since the last term in  \eqref{eq:HCGmu} just contributes to the chemical potential term in the grand-canonical Hamiltonian \eqref{eq:K}, we are free to  write  
\begin{align}
   \hat{H}^{CG}_{I} &= \frac{V_0}{\mathcal{V}} N_T^2
\end{align}
and update $\mu$ as $\mu \rightarrow \mu - V_0 n$.  Replacing $V_0$  with $g^{G}_0$ then results in \eqref{eq:HCG}. 

Since the classical gravity interaction Hamiltonian only affects the ground state energy, we only need the usual Bogoloybov transformation that is used to  diagonalize a BEC with electromagnetic interactions \cite{PitaevskiiBook}, which is as \eqref{eq:uk}-\eqref{eq:ekQGapp} with $g_k^{G}$ set to zero.

\subsection{Phonons and interactions} \label{app:NGBosons}

The Nambu-Goldstone theorem connects the breaking of a global internal symmetry of a system to the onset of vector-bosons of either odd or even power in the momentum variable $\bm{k}$, and in that sense it holds for a wide variety of fields from high-energy to condensed matter physics.

A discussion of the connection between Bose-Einstein condensation and the breaking of the $U (1)$-symmetry of the system at the second order phase transition can be found, for instance, in \cite{LIEB2007389}. Today, the categorization is often given as Type-A and Type-B, as for instance in \cite{2020ARCMP..11..169W}, where counting rules for NGBs are discussed. For a very recent and extensive pedagogical derivation of both the \emph{Nambu-Goldstone Theorem} as well as the classification of Nambu-Goldstone-bosons, see \cite{2024arXiv240414518B}.

The quadratic dispersion relation for a quantized gravity picture corresponds to what has been categorized first in \cite{NIELSEN1976445} as \emph{Type-II Nambu-Goldstone bosons} (NGBs).

When going beyond the contact approximation of the electromagnetic interactions, it is likely that phonons (Type-I NGBs) no longer necessarily emerge at low momenta in a homogeneous, isotropic BEC. However, the observation of phonons in experiments evidences that the contact approximation must be a very good approximation for the electromagnetic interactions.

\subsection{Validity of approximations} \label{app:Approxs}

Given the much higher mass required to evidence quantum gravity using the heat capacity or energy spectrum of the BEC compared to current experiments, we should make sure that the approximations used to derive the results are still valid. For example, with a high mass, we should check that assuming a non-relativistic BEC and the Newtonian limit of gravity are adequate approximations in deriving our results.  This requires that the Newtonian gravitational self-energy of the BEC is much less than its rest energy. Approximating the BEC as spherical and of uniform density for simplicity, this requires:
\begin{align}
    \frac{3 G M^2}{5 R} &\ll M c^2\\
    \implies R &\ll \frac{ 3 G M}{5 c^2},
\end{align}
which is approximately the Schwarzschild radius of the BEC, $R_s = G M / c^2$. We might expect that relativistic effects become important when $R \approx 10^9 \, R_s$ \cite{thorne2000gravitation}, which is the case for the Earth. With $N = 10^{15} - 10^{16}$ atoms of $^{174}{\mathrm{Yb}}$, the radius would need to be smaller than $10^{-25}\,\mathrm{cm}$ for relativistic effects to be relevant, resulting in a BEC of mass density $\rho \approx 10^{69} - 10^{70}\, \mathrm{kg \,m^{-3}}$.  Instead, we assume $R \approx 1\,\mathrm{cm}$ in the main text, resulting in a  mass density of $10^{-3} - 10^{-4}\,\mathrm{kg \, m^{-3}}$ for the BEC. We are, therefore, far from the relativistic regime of gravity and approximating a non-relativistic BEC as in \eqref{eq:K} with Newtonian gravity is well-justified.

Another approximation we make in deriving the heat capacity of a self-gravitating BEC, is to use Bogoliubov perturbation theory where we neglect terms in the Hamiltonian involving fewer than two $\hat{\phi}(\bm{r})$ ground-state operators.  This approximation is actually improved with greater numbers of condensed atoms and lower temperatures, as it requires $N_0 \gg 1$ and $N_0 \approx N$ \cite{PitaevskiiBook}. Another  approximation that we implicitly made in using the Bogoliubov approximation is that the BEC is dilute: $n |a|^3 \ll 1$. Despite the high mass of the required BECs, since the volume is also relatively large ($L = 1\,\mathrm{cm}$),  the number density, $n = 10^{21} - 10^{22}\,\mathrm{m^{-3}}$, is only about an order of magnitude greater than the most dense BECs in current experiments. The diluteness condition is thus also satisfied by the BECs we require to evidence quantum gravity: with $a \approx 1\,\mathrm{nm}$ and $n = 10^{22} \,\mathrm{m^{-3}}$, then $n a^3 \approx 10^{-4}$. At these densities, the atoms will be moving non-relativistically, which can be demonstrated using the Heisenberg uncertainty principle $\Delta x \, \Delta p \geq \hbar/2$. Taking $\Delta p \sim m v$, with $v$ the velocity of the atoms, and $\Delta x \sim n^{-1/3}$, then $ v \sim 4\,\mathrm{mm\,s^{-1}}$, which is a typical value for the speed of sound of a BEC and far from a relativistic regime.

\subsection{Mean field potential}

Fundamental semi-classical gravity \cite{ROSENFELD1963353,moller1962theories} in the first-quantized  Newtonian regime for a single system is described by the (fundamental)  Schr\"{o}dinger-Newton equation \cite{1984PhLA..105..199D}:
\begin{align}
    i \hbar \frac{\partial \psi(t,\bm{r})}{\partial t} =  -\frac{\hbar^2}{2m} \nabla^2 \psi(t,\bm{r}) + V_{ext}(\mathbf{r}) \psi(t,\bm{x})  - G m^2\, \psi(t,\bm{r}) \int \frac{|\psi(t, \mathbf{r}')|^2}{|\mathbf{r} - \mathbf{r}'|} \, \mathrm{d}^3 r',
\end{align}
where $\psi(t,\bm{r})$ is the wavefunction of the particle, $m$ is its mass, and $V_{ext}$ is an external potential. In the main article, we considered fundamental semi-classical gravity in the Newtonian regime but second-quantized picture, such that we do not work explicitly with the fundamental Schr\"{o}dinger-Newton equation. This equation, however, is mathematically similar to a well-used equation in Bose gasses, the time-dependent Gross-Pitaevskii equation \cite{gross1961structure,pitaevskii1961vortex}:
\begin{align}
    i \hbar \frac{\partial \phi(t,\mathbf{r})}{\partial t} =  -\frac{\hbar^2}{2m} \nabla^2  \phi(t,\mathbf{r}) + V_{ext}(\mathbf{r}) \phi(t,\mathbf{r}) + g^{EM} \,\phi(t,\mathbf{r})|\phi(t,\mathbf{r})|^2,
\end{align}
where $\phi(t,\mathbf{r})$ is the classical field, often refereed to as the wavefunction, of the condensate, and $g|\phi(t,\mathbf{r})|^2$ is often referred to as the mean field potential since it is the `average' potential a  particle experiences due to all others. The only difference between the two equations, mathematically, is that the the electromagnetic interaction is taken to be a contact interaction in the latter, as discussed in the main article.

The time-dependent Gross-Pitaevskii equation  can be considered as the zeroth order equation coming from the Bogoliubov method. As stated in the main text, in this method we approximate $\hat{\Psi}(\bm{r}) = \phi(\bm{r}) + \hat{\psi}(\bm{r})$, and then plug this into \eqref{eq:K}, after which we keep only terms quadratic in $\hat{\psi}(\bm{r})$. At zeroth order, we neglect the quantum fluctuations  $\hat{\psi}(\bm{r})$ entirely, such that the canonical Hamiltonian \eqref{eq:K} only contains the classical field $\phi(\bm{r})$ for the  condensate. Adding an external potential, the  variation of this  with respect to $\phi^{\ast}(\bm{r})$ then gives the time-dependent Gross–Pitaevskii equation, which describes the evolution of the condensate in the assumption that the quantum fluctuations can be neglected. With $\langle \hat{\psi}(\bm{r})\rangle = 0$, then $\phi(\bm{r}) = \langle \hat{\Psi}(\bm{r})\rangle$, such that $\phi(\bm{r})$ can also be considered a `mean' field.

The time-dependent Gross–Pitaevskii equation  is not used in the main article  as $V_{ext}=0$ is assumed with periodic boundary conditions, such that the condensate is always uniform if it is uniform to begin with \cite{PitaevskiiBook}, and the heat capacity only depends on the energy spectrum of the quasi-particles \cite{PitaevskiiBook}. This is the case irrespective of whether gravity is quantum or classical, and in fact the  time-dependent Gross–Pitaevskii equation is formally the same in both cases since  the interaction Hamiltonians for quantum gravity and classical gravity \eqref{eq:HIG} and \eqref{eq:HICG} are formally the same in the approximation $\hat{\Psi}(\bm{r}) = \phi(\bm{r})$. The assumption of a uniform BEC is made predominately to simplify calculations, following standard treatments of electromagnetically interacting BECs, see for example Chapter 4 of \cite{PitaevskiiBook}.  Creating and maintaining a uniform density in experiments has been achieved \cite{UniformBECExp,PhysRevLett.118.210401,PhysRevLett.119.190404,PhysRevLett.119.250404}, but a non-uniform BEC for the gravity experiment is likely more realistic. Although the heat capacity does not count over the energy of the ground state \eqref{eq:cv}, it implicitly depends on the density of the condensate via the full Bose gas density $n$ in the quasi-particle energies \eqref{eq:EigenEnerQ} and \eqref{eq:ekCG1}. Therefore the heat capacity of a non-uniform BEC will, in general, be different to that of a uniform one. However, note that the density is always accompanied by the coupling constant of gravity or electromagnetism in \eqref{eq:EigenEnerQ} and \eqref{eq:ekCG1}, which is likely to also be the case for a non-uniform BEC given electromagnetic analogues \cite{PitaevskiiBook}. Therefore, if gravity were to weakly perturbed the condensate density, this effect would contribute a second order dependence in $g_0^G$ for the heat capacity and can thus be neglected. This is the case for the experimental values considered in the main text:  with $n = 10^{22} \,\mathrm{m^{-3}}$, $L = 1\,\mathrm{cm}$ and Feshbach resonances making sure $g^G_0 \approx g^{EM}$, the mean field component of the Gross-Pitaevskii equation is much weaker than the kinetic part since $\hbar^2 / (2m L^2) \gg g^G_0 n$. Therefore, even for a non-uniform BEC, the effect of gravity on the condensate density will have a negligible contribution to the heat capacity, as long as the kinetic term is stronger than gravity's contribution to the mean field, which is likely for realistic BECs.

In general, assuming a non-uniform BEC will likely result in a different numerical change in the heat capacity due to quantum and classical gravity compared to a uniform BEC. However,  there is no reason why the numerical difference between the heat capacity with quantum and classical gravity would become unmeasurable when a non-uniform BEC is considered compared to a uniform one, particularly as the Gross-Pitaevskii equations are the same for quantum  and classical gravity as stated above. Our main idea is, therefore, independent of the uniformity of the BEC or how the density of the BEC would change with gravity.

Despite the mathematical similarity, the two equations, the fundamental Schr\"{o}dinger-Newton equation and the Gross–Pitaevskii equation, describe very different physical phenomena. The fundamental Schr\"{o}dinger-Newton equation  is based on the idea that gravity is fundamentally classical, and thus applies even to a single particle, where  its own wavefunction self-gravitates \cite{Bahrami_2014}. In comparison, the Gross–Pitaevskii equation only applies to a many-body system.  For a more detailed discussion on the connection and differences between these two equations, see e.g.\ \cite{PAREDES2020132301}.

\subsection{Derivation of BEC Hamiltonian with quantum gravity}

Here, we derive the interaction Hamiltonian \eqref{eq:HIG} from first-principles. Effective field theory tells us that any correct theory of quantum gravity must approximate  perturbative quantum gravity at low energies \cite{Carney_2019}. In this theory, gravity is assumed weak so that, assuming a flat background, we can split the metric as $g_{\mu \nu}  = \eta_{\mu \nu} + h_{\mu \nu}$ with $|h_{\mu \nu}| \ll 1$. Then, only the perturbed metric is quantized: $h_{\mu \nu} \rightarrow \hat{h}_{\mu \nu}$. The action for gravity coupled to matter in this regime is \cite{maggiore2008gravitational}:
\begin{align}
    S = \frac{c^4}{64 \pi G} \int \mathrm{d}^4 x \left( 
    - \partial_\rho \hat{h}_{\mu\nu} \partial^\rho \hat{h}^{\mu\nu}
    + 2 \partial_\rho \hat{h}_{\mu\nu} \partial^\nu \hat{h}^{\mu\rho}
    - 2 \partial_\mu \hat{h}^{\mu\nu} \partial_\nu \hat{h}
    + \partial^\mu \hat{h} \partial_\mu \hat{h}
\right)
+ \frac{1}{2} \int \mathrm{d}^4 x \, \hat{h}_{\mu\nu} \hat{T}^{\mu\nu},
\end{align}
where $\hat{T}_{\mu \nu}$ is the stress-energy tensor for matter. If gravity is only sourced by the matter described by $\hat{T}^{\mu\nu}$, then the above action simplifies to \cite{Christodoulou:2022mkf}:
\begin{align}
S = \frac{1}{4} \int d^4 x \, \hat{h}_{\mu \nu} \hat{T}^{\mu \nu},
\end{align}
such that the interaction Hamiltonian between matter and gravity is:
\begin{align}
    \hat{H}_{int} = - \frac{1}{4} \int d^3 \bm{r}\,\hat{h}_{\mu \nu} \hat{T}^{\mu \nu}.
\end{align}
In the Newtonian regime, $h_{\mu \nu}(\bm{r}) \approx - 2 \Phi(\bm{r}) \delta_{\mu \nu} / c^2$ \cite{poisson2014gravity}, where $\Phi(\bm{r})$ is the Newtonian potential. Then, assuming non-relativistic matter, such that $T_{00} = \rho c^2$ dominates over all other components, with $\rho$ the mass density, we obtain:
\begin{align}
    \hat{H}_{int} =  \frac{1}{2} \int d^3 \bm{r} \,\hat{\Phi} (\bm{r}) \hat{\rho}(\bm{r}).
\end{align}
As stated above, gravity is only sourced by the considered matter, and thus:
\begin{align} \label{eq:hatPhi}
    \hat{\Phi} (\bm{r}) = - G \int d^3 \bm{r}' \,\frac{\hat{\rho}(\bm{r}')}{|\bm{r} - \bm{r}'|},
\end{align}
such that:
\begin{align} \label{eq:HintQG}
    \hat{H}_{int} =  -\frac{1}{2} G  \int d^3 \bm{r}\,  d^3 \bm{r}'\, \frac{\hat{\rho}(\bm{r}) \hat{\rho}(\bm{r}')} {|\bm{r} - \bm{r}'|}.
\end{align}
This Hamiltonian could have also been derived straight from the Hamiltonian theory of Newtonian gravity and quantizing matter \cite{PRXQuantum.2.010325}. In a field description of a BEC, the mass density operator is $\hat{\rho}(\bm{r}) = 
 m \hat{\Psi}^{\dagger}(\bm{r}) \hat{\Psi}(\bm{r}) $ \cite{PitaevskiiBook}, where $m$ is the mass of the atoms, which results in \eqref{eq:HIG} ignoring the infinite term resulting from $[\hat{\Psi} (\bm{x}), \hat{\Psi}^{\dagger}(\bm{x}')] = \delta^{(3)}(\bm{x} - \bm{x}')$ under the standard quantum field assumption that only differences in energy are physical.

 Above we have assumed a quantum description of gravity. However, so far there is no strong experimental evidence that the gravitational interaction is quantized, and it has been suggested that gravity could be fundamentally classical \cite{ROSENFELD1963353,moller1962theories,1984PhLA..105..199D,penrose1996gravity,Kafri_2014,OppenheimCG,TilloyDiosi}. Our experimental proposal, as with other recently proposed table-top experiments (see e.g.\ \cite{bose2023massivequantumsystemsinterfaces}), aims to  evidence the quantum nature of gravity. While the grand-canonical Hamiltonian \eqref{eq:K} with \eqref{eq:HIG} describes a BEC self-interacting through a quantum version of gravity in the Newtonian regime, it cannot describe a BEC interacting with a classical gravity interaction. For example, we can consider the above derivation but keeping gravity classical. In this case, we would keep to $g_{\mu \nu} = \eta_{\mu \nu} + h_{\mu \nu}$ rather than quantizing $h_{\mu \nu}$. The interaction Hamiltonian between quantum matter and classical gravity, before we consider how  gravity is sourced, is then
 \begin{align}
 \hat{H}_{int} = -\frac{1}{2} \int d^3 \bm{x} h_{\mu \nu} \hat{T}_{\mu \nu}.     
 \end{align}
 In a quantum version of gravity, we have, in general before fixing a gauge:
 \begin{align}
G^{(1)}_{\mu\nu}[\hat{h}] = -\frac{16\pi G}{c^4} \hat{T}_{\mu\nu}
 \end{align}
where the linearized Einstein tensor \( G^{(1)}_{\mu\nu} \) is defined as \cite{maggiore2008gravitational}:
\begin{align} \label{eq:G1}
G^{(1)}_{\mu\nu}[h] =\Box \bar{h}_{\mu \nu} + \eta_{\mu \nu} \partial^{\rho} \partial^{\sigma} \bar{h}_{\rho \sigma} - \partial^{\rho} \partial_{\nu} \bar{h}_{\mu \rho} - \partial^{\rho} \partial_{\mu} \bar{h}_{\nu \rho} 
\end{align}
with $ \bar{h}_{\mu \nu} := h_{\mu \nu} - \eta_{\mu \nu} h$ and $h :=  \eta^{\mu \nu} h_{\mu \nu} $. In the Lorentz gauge, this simplifies to:
\begin{align}\label{eq:lorentz}
    \Box \hat{\bar{h}}_{\mu \nu} = - \frac{16 \pi G}{c^4} \hat{T}_{\mu \nu}.
\end{align}
 However, in our  classical theory of gravity, while $\hat{T}_{\mu \nu}$ is a quantum operator, $h_{\mu \nu}$ is not, and so \eqref{eq:G1} or \eqref{eq:lorentz} cannot be satisfied since we would have a quantum operator on one side but not the other.  One solution, traditional semi-classical gravity, is to take the expectation value of the matter quantum operator:
 \begin{align} \label{eq:G1classical}
     G^{(1)}_{\mu\nu}[h] = -\frac{16\pi G}{c^4} \langle \hat{T}_{\mu\nu} \rangle,
 \end{align}
 but there are also other options, such as stochastic fluctuations around this \cite{TilloyDiosi,layton2023weak}. As stated above, in the non-relativistic limit, $h_{\mu \nu}(\bm{r}) = - 2 \Phi(\bm{r}) \delta_{\mu \nu} / c^2$, and we just have $\hat{T}_{00} = \hat{\rho} c^2$, such that the interaction Hamiltonian becomes:
  \begin{align} \label{eq:HintClassical}
 \hat{H}_{int} = \int d^3 \bm{r}\, \Phi (\bm{r}) \hat{\rho}(\bm{r}),  
 \end{align}
 with \eqref{eq:G1classical} giving 
\begin{align} \label{eq:PhiClassical}
    \Phi (\bm{r}) = - G \int d^3 \bm{r}' \,\frac{\langle \hat{\rho}  (\bm{r}')\rangle}{|\bm{r} - \bm{r}'|}.
\end{align}
Plugging \eqref{eq:PhiClassical} into \eqref{eq:HintClassical}, we obtain  
 \begin{align}
    \hat{H}_{int} =  - G  \int d^3 \bm{r}\,  d^3 \bm{r}'\, \frac{\langle \hat{\rho}(\bm{r'})\rangle \hat{\rho}(\bm{r})} {|\bm{r} - \bm{r}'|}.
\end{align}
With $\hat{\rho} = m \hat{\Psi}^{\dagger} \hat{\Psi}$, we then obtain \eqref{eq:HICG}. Note that this is not the same as \eqref{eq:HIG}. As with quantum gravity, we can also obtain this directly from the Hamiltonian of the fully classical Newtonian theory:
  \begin{align} \label{eq:HintClassical}
 \hat{H}_{int} = \int d^3 \bm{r}\, \Phi (\bm{r}) \rho(\bm{r}),     
 \end{align}
and then quantizing $\rho$ and using \eqref{eq:PhiClassical} for how the classical potential $\Phi$ can be sourced by quantum matter.

It has been argued that \eqref{eq:HintQG} can be, contrary to what we described above, an interaction Hamiltonian of a classical theory of gravity  \cite{anastopoulos2018comment,anastopoulos2021gravitational}. This is because the fully Newtonian theory of classical gravity (without quantum matter), and linear classical gravity in the Coulomb gauge (without quantum matter), contain the Hamiltonian \cite{Christodoulou_2023}:
 \begin{align}
    H_{int} =  -\frac{1}{2} G  \int d^3 \bm{r}\,  d^3 \bm{r}'\, \frac{\rho(\bm{r'}) \rho(\bm{r})} {|\bm{r} - \bm{r}'|}.
\end{align}
Then we might consider that, with  matter quantized, we just end up with \eqref{eq:HintQG} even in a classical theory of gravity, since we just take $\rho \rightarrow \hat{\rho}$. However, this misses the fact that the potential would then be defined as \eqref{eq:hatPhi}, such that gravity is quantum - the gravitational potential is fully defined by matter, which has been taken to obey quantum theory (the quantized potential then acts on the Hilbert space of quantum matter \cite{Christodoulou_2023}). This is not possible in a classical theory of gravity, as illustrated above. For example, a quantum superposition of matter states would naturally lead to a quantum superposition of Newtonian gravitational forces given \eqref{eq:HintQG}, which should not be possible in a classical theory of gravity, and indeed would not occur with  \eqref{eq:PhiClassical}.

\end{document}